\documentclass[aps,prb,showpacs,preprintnumbers]{revtex4}
\usepackage{graphicx}
\begin{document}
\preprint{Cond-mat/0208543; Published in JPCM 15,4411(2003).}
\title{Theory of nonlinear cyclotron resonance
in quasi-two-dimensional electron systems}
\author{S.Y. Liu}
\email{liusy@mail.sjtu.edu.cn}
\author{X.L. Lei}
\affiliation{Department of Physics, Shanghai Jiaotong University,
1954 Huashan Rd., Shanghai 200030, China}

\begin{abstract}

Momentum and energy balance equations are developed for steady-state electron
transport and optical absorption under the influence of a dc electric field, 
an intense ac electric field of terahertz (THz) frequency in a two-dimensional 
(2D) semiconductor in the presence of a strong magnetic field perpendicular to 
the 2D plane. These equations are applied to study the intensity-dependent
cyclotron resonance (CR) in far-infrared transmission and THz-radiation-induced 
photoconductivity of GaAs heterostructures in Faraday geometry.
We find that the CR peaks and line shapes of the transmittance exhibit different 
intensity dependence when the intensity of THz field increases in the range above 
or below a certain critical value.
The CR in photoresistivity, however, always enhances with increasing the
intensity of the THz field. These results qualitatively agree with the
experimental observations. We have clarified that the CR in photoconductivity
is not only the result of the electron heating, but also comes from photon-assisted 
scattering enhancement, especially at high temperatures. 
The effects of an intense THz field on Faraday angle and
ellipticity of magnetically-biased 2D semiconductors have also been
demonstrated.
 
\end{abstract}

\pacs{73.50.Jt, 73.50.Mx, 78.67.De, 78.20.Ls}

\maketitle
\section{Introduction}
Cyclotron resonance (CR) is a fundamental process of carriers in quasi-two-dimensional 
semiconductors subjected simultaneously to a magnetic field and a far-infrared
or terahertz (THz) ac field. It occurs when the separation between two adjacent
Landau levels is closed to the photon energy of the ac field, and leads to the resonant
behavior in absorption and in photoconductivity as functions of the magnetic field.

CR in absorption or transmission has been proved to be a powerful tool providing valuable
information of two-dimensional (2D) semiconductors on the carrier scattering processes, 
the confining potential, and even the spin relaxation mechanisms.\cite{Kono} 
Generally, weak far-infrared irradiations were used in most of experiments, and  
theoretical analyses focused on the linear response of the system to the ac 
field.\cite{Ting}  
When the strength of the incident radiation field increased, nonlinear behavior of CR in
absorption and transmission was observed. It was found\cite{Rodriguez} that the resonant field to
zero field transmissivity ratio descended slightly with increasing the intensity of the
far-infrared field from 0.1\,W/cm$^2$, then increased for radiation-field intensities
above 10\,W/cm$^2$. This kind of intensity-dependent CR transmission
has not yet been explained so far except a simple comparison with linear-response 
formula of Drude-type taking the scattering time as a fitting parameter. 

Another manifestation of CR shows up in photoresponse, namely, photoconductivity or 
photoresistivity. The latter is defined as the longitudinal dc 
magnetoresistivity change induced by the irradiation of the high-frequency electromagnetic
wave, which exhibits a resonant peak structure under CR condition. 
This effect has long been known at low temperatures, and was believed
to arise from the electron heating induced redistribution of
the electron gas upon absorbing the radiation-field energy.\cite{CRP,New}
The energy absorption and thus the rise of the electron temperature exhibit
maxima when the photon energy of the radiation field equals the
cyclotron energy, yielding the resonance peak structure of the photoresistivity.
Later analysis of experimental data \cite{Mordovets} indicated that,
in addition to this heating-induced electron redistribution, another nonthermal 
mechanism might be needed to account for the observed photoconductivity of GaAs/AlGaAs
heterostructures at CR. Recent photoresistivity measurement of GaAs/AlGaAs heterojunction 
performing at lattice temperature $T=150\,$K  subjected to irradiations of 4\,THz frequency,
\cite{Koenraad} surprisingly showed remarkable CR peaks. At such a high lattice temperature
with strong polar optic phonon scattering providing an efficient energy dissipation 
channel, the radiation-induced electron-temperature increase is far smaller to account for
such strong CR in photoresistivity.  
So far no convincing nonthermal mechanism responsible for far-infrared photoconductivity
in semiconductors has been proposed. 
A nonlinear theory for high-temperature photoresistivity CR
in two-dimensional system is still lacking.  
It is an urgent need to develop a tractable, microscopic theory capable of treating 
nonlinear absorption and photoresponse under intense THz radiation and to clarify the 
missing nonthermal mechanism for high-temperature photoresistivity 
in two-dimensional semiconductors.
 
A few years ago, one of the authors developed a balance-equation approach \cite{Lei}
for hot-electron transport driven by a THz electric field of single frequency 
(${\bf E}_{\omega}\sin(\omega t)$). This method made use of the fact that,
when the harmonic generation is small and the frequency gets 
into the THz regime or higher, the electron drift velocity
in the steady transport state oscillates almost out of phase 
of the electric field, i.e. the drift velocity is essentially of the form  
${\bf v}_1\cos(\omega t)$. At the same time, all orders of this (frequency $\omega$) photon
assisted impurity and phonon scatterings are included in the relaxation processes.
This method has been successfully applied to discuss
THz photoabsorption and THz-induced dc conductivity response in bulk and two-dimensional
semiconductors in the case without magnetic field\cite{Lei,Lei00} or with a magnetic field in Voigt
configuration.\cite{Lei00E} However, this assumption of such time-dependent form of 
drift velocity will no longer be true when there is a 
strong magnetic field ${\bf B}$ not parallel to ${\bf v}_1$.  
Since the Lorentz force $e{\bf v}_1\times{\bf B}\cos(\omega t)$ acts on
the moving electron, its steady-state drift velocity will contain a term 
$(\omega_c/\omega)v_1\sin(\omega t)$,
where $\omega_c=|eB|/m$ is the cyclotron frequency.
This velocity, which is perpendicular to ${\bf v}_1$ and ${\bf B}$, and oscillates 
$\pi/2$ out of phase of ${\bf v}_1\cos(\omega t)$,
is of the same order of magnitude as ${\bf v}_1$ in the frequency and magnetic field
range ${\omega}\sim{\omega}_c$. Because of this, the balance-equation
method used in Ref.\,\onlinecite{Lei00E} is not able to deal with the problem related to cyclotron resonance in semiconductors in Faraday geometry.

The purpose of this article is to develop a theory for electron
transport driven by an intense far-infrared electric field 
in a quasi-2D semiconductor in the presence of 
an arbitrary magnetic field perpendicular to the 2D plane.  
We find that it is possible to extend balance equation method proposed in
Ref.\,\onlinecite{Lei} to the case in the presence of a quantized magnetic
field in Faraday configuration, 
and to deal with those problems in magnetotransport, such as
the intensity-dependent THz transmission and photoresponse at CR
in quasi-2D electron systems.

This paper is organized as follows. In Sec.\,II we will sketch the derivation
of the force- and energy-balance equations of quasi-two-dimensional
electron systems subjected to a magnetic field,
an arbitrary dc and THz ac electric fields.
In Sec.\,III the cyclotron resonance in
time-dependent drift velocity is discussed.
The investigation on cyclotron resonance in transmission
and photoconductivity is, respectively,
presented in Sec.\,IV and Sec.\,V.
Finally, in Section VI a short conclusion will be given.
 
\section{FORMULATION}
\subsection{Hamiltonian}
We consider $N_{\rm e}$ electrons in a unit area of a quasi-two-dimensional system, such as
a heterojunction or a quantum well. In these systems the electrons are free to move in the
$x$-$y$ plane, but are subjected to a confining potential $V(z)$ in the
$z$-direction. These electrons are interacting with each other and also coupled
with phonons and scattered by randomly distributed impurities in the lattice. 

To include possible elliptically polarized electromagnetic radiation we assume that 
a uniform dc electric field ${\bf E}_0$ and  
a terahertz ac field ${\bf E}(t)$ of angular frequency $\omega$,
\begin{equation}
{\bf E}(t)\equiv{\bf E}_s \sin(\omega t)+{\bf E}_c\cos(\omega t),
\end{equation}
are applied in the $x$-$y$ plane, together with a uniform magnetic field ${\bf B}=(0,0,B)$ 
along the $z$ axis.
These magnetic and electric fields can respectively be described by
a vector potential ${\bf A}({\bf r})$
and a scalar potential $\varphi({\bf r},t)$ of the form
\begin{equation}
{\bf \nabla}\times {\bf A}({\bf r})={\bf B},
\end{equation}
\begin{equation}
\varphi({\bf r},t)=-{\bf r}\cdot {\bf E}_0-{\bf r}\cdot{\bf E}(t).
\end{equation}
In the presence of these electric and magnetic fields the Hamiltonian
of the system has the form
\begin{equation}
H=H_{\rm eE}(t)+H_{\rm ei}+H_{\rm ep}+H_{\rm ph}.
\end{equation}
Here
\begin{equation}
H_{\rm eE}(t)=\sum_{j}\left[\frac{1}{2m}
\left({\bf p}_{j\|}-e{\bf A}({\bf r}_{j\|})\right)^{2}
+\varphi({\bf r}_{j\|},t)+\frac{p_{jz}^2}{2m_z}+V(z_j)\right]+\sum_{i<j}V_c({\bf r}_{i\|}-{\bf r}_{j\|},z_i,z_j)\label{eqHeE}
\end{equation}
is the Hamiltonian of electrons driven by the electric
and magnetic fields with $V_c$ being the electron-electron
Coulomb interaction, $H_{\rm ei}$ and $H_{\rm ep}$ are, respectively,
the electron-impurity and electron-phonon couplings. In equation
(\ref{eqHeE}), ${\bf r}_{j\|}\equiv (x_j,y_j)$ and
${\bf p}_{j\|}\equiv(p_{jx},p_{jy})$ are the coordinate and momentum of
the $j$th electron in the 2D plane,
and $z$ and $p_{jz}$ are those perpendicular to the plane; $m$ and $m_z$ are,
respectively, the effective mass
parallel and perpendicular to the plane.

It is convenient to introduce two-dimensional center-of-mass momentum
and coordinate variables
${\bf P}\equiv(P_x,P_y)$ and ${\bf R}\equiv(R_x,R_y)$:
\begin{equation}
{\bf P}=\sum_j {\bf p}_{j\|},\,\,\,\,\,\,\,\,\,\,{\bf R}=\frac 1{N_{\rm e}}\sum_j {\bf r}_{j\|}
\end{equation}
and the relative-electron momentum and coordinate variables
${\bf p}'_{j}\equiv({\bf p}'_{j\|},p_{jz})$ and
Š${\bf r}'=({\bf
r}'_{j\|},z_j)$($j=1,...N_{\rm e}$):
\begin{equation}
{\bf p}_{j\|}'={\bf p}_{j\|}-\frac 1{N_{\rm e}}{\bf P},\,\,\,\,\,\,\,\,\,\,{\bf r}_{j\|}'={\bf r}_{j\|}-{\bf R}.
\end{equation}
In term of these variables, the Hamiltonian $H_{\rm eE}$ can be separated
into a center-of-mass part $H_{\rm cm}$
and a relative electron part $H_{\rm er}$:
\begin{eqnarray}
H_{\rm eE}&=&H_{\rm cm}+H_{\rm er},\\
H_{\rm cm}&=&\frac 1{2N_{\rm e}m}({\bf P}-N_{\rm e}e{\bf A}({\bf
R}))^2-N_{\rm e}e{\bf E}_{0}\cdot {\bf R}-N_{\rm e}e{\bf E}(t)\cdot {\bf R},\\
H_{\rm er}&=&\sum_{j}\left[\frac{1}{2m}\left({\bf p}_{j\|}'-e{\bf A}({\bf r}_{j\|}')\right)^{2}
+\frac{p_{jz}^2}{2m_z}+V(z_j)\right]+\sum_{i<j}V_c({\bf r}_{i\|}'-{\bf r}_{j\|}',z_i,z_j).
\end{eqnarray}
It should be noted that the relative-electron Hamiltonian $H_{\rm er}$ is the just 
that of a quasi-2D system subjected to a uniform magnetic field in the $z$ direction. 
Its eigenstate can be designated by a subband
index $s$, a Landau level index $n$ and a wavevector $k_y$, having energy spectrum
\begin{equation}
\varepsilon_{sn}=\varepsilon_s+(n-\frac 12)\omega_c,\,\,\,\,\,s=0,1,...\,\,\,\,{\rm and}
\,\,\,\,n=1,2,...
\end{equation}
where $\omega_c=|eB|/m$ is the cyclotron frequency. In this paper we will limit to the
case that the 2D electrons
occupying only the lowest subband and ignore index $s$.
In Eq.\,(5) the electron-impurity and electron-phonon interaction
$H_{\rm ei}$ and $H_{\rm ep}$ have the same expression as those given
in Ref.\,\onlinecite{Lei851}, in terms of
center-of-mass coordinate ${\bf R}$ and the density operator of the relative-electrons,
\begin{equation}
\rho_{{\bf q}_{\|}}=\sum_j {\rm e}^{{\rm i}{\bf q}_{\|}\cdot {\bf r}_j'}.
\end{equation}

\subsection{Balance equations}
On the basis of the Heisenberg equation of motion we can derive the
velocity (operator) of the center-of-mass, $\bf V$, which is the
rate of change of the center-of-mass coordinate ${\bf R}$, and
equation for the rate of change of the center-of-mass velocity
${\bf V}\equiv \dot{{\bf R}}$:
\begin{equation}
{\bf V}=-{\rm i}[{\bf R},H]=\frac {1}{N_{\rm e}m}({\bf P}-N_{\rm e}e{\bf
A} ({\bf R})),
\end{equation}
and
\begin{equation}
\dot{{\bf V}}=-{\rm i}[{\bf V},H]+\frac {\partial {\bf V}}{\partial t}=
\frac{e}{m} \left\{{\bf E}_0+{\bf E}(t)+
{\bf V} \times {\bf B}\right\}+
\frac {\bf F}{N_{\rm e}m},\label{eqdotv}
\end{equation}
with
\begin{equation}
{\bf F}=-{\rm i}\sum_{{\bf q}_\|,a}U({\bf q}_\|,z_a){\bf q}_\|{\rm e}^
{{\rm i}{\bf q}_\|\cdot ({\bf R}-{\bf r}_a)}\rho_{{\bf q}_\|}-
{\rm i}\sum_{{\bf q},\lambda}M({\bf q},\lambda)\phi_{{\bf q}\lambda}
{\rm e}^{{\rm i}{\bf q}_\|\cdot {\bf R}}\rho_{{\bf q}_\|}.
\label{eqopf}
\end{equation}
We can also derive equation for the rate of change of the relative
electron energy:
\begin{equation}
\dot{H}_{\rm er}=-{\rm i}[H_{\rm er},H]=-{\rm i}\sum_{{\bf q}_\|,a}U({\bf q}_\|,z_a)
{\rm e}^{{\rm i}{\bf q}_\|\cdot ({\bf R}-{\bf r}_a)}\dot{\rho}_{{\bf
q}_\|}-{\rm i}\sum_{{\bf q},\lambda}M({\bf q},\lambda)
\phi_{{\bf q}\lambda}
{\rm e}^{{\rm i}{\bf q}_\|\cdot {\bf R}}\dot{\rho}_{{\bf
q}_\|}.\label{eqopdh}
\end{equation}
In Eqs.\,(\ref{eqopf},\ref{eqopdh}) (${\bf r}_a$,$z_a$) and $U({\bf
q}_\|,z_a)$ are the impurity position and its potential, $M({\bf q},
\lambda)$ is the matrix element due to coupling between electrons and a
phonon of wave vector ${\bf q}\equiv ({\bf q}_\|,q_z)$ in branch $\lambda$ having energy
$\Omega_{{\bf q}\lambda}$, $\phi_{{\bf q}\lambda}\equiv b_{{\bf q}
\lambda}+b^{\dagger}_{-{\bf q}\lambda}$ stands for the phonon field operator,
and $\dot{\rho}_{{\bf q}_\|}\equiv -{\rm i}[\rho_{{\bf q}_\|},H_{\rm
er}]$.

In order to derive the force- and energy-balance
equations we need to carry out the statistical average of these operator equations.
As proposed in Ref.\,\onlinecite{Lei85}, we treat
the center-of-mass coordinate ${\bf R}$ and velocity ${\bf V}$ classically, and
by neglecting their small fluctuations we will regard them as time-dependent
expectation values of the center-of-mass coordinate and velocity,
${\bf R}(t)$ and ${\bf V}(t)$.
In the present paper, we are mainly concerned with the steady transport state under 
a terahertz irradiation of single frequency and focus on the photoresponse of 
the dc conductance and the energy absorption and transmission of the THz signal. 
These quantities are directly related to the time-averaged and base-frequency 
oscillating components of the center-of-mass velocity.
The second and higher harmonic components of the electron velocity, if any, 
though give no direct contribution to ac-field transmission and dc photoresponse, 
would enter the frictional force, energy transfer and energy absorption rates 
in the resulting equations, thus may in turn affect the time-averaged and 
lower harmonic terms of the drift velocity.
However, unless for a specially designed system, the generated power of the third harmonic 
current in an ordinary semiconductor is generally less than a few percent of the 
base-frequency power even in the case of strongly nonlinear transport when the ac field
amplitude gets as high as several kV/cm or higher.\cite{Lei97,Mayer} 
Its effect on the photon-assisted scattering matrix element will be an order of 
magnitude further weaker than that of the base frequency photons.\cite{Lei,Lei00,Lei00E} 
For the radiation field intensity concerned in the present study, which is an order 
of magnitude smaller than that for the above-mentioned harmonic generation, the effect 
of higher harmonic current is safely negligible. Hence,
it suffices to assume that the center-of-mass 
velocity, i.e. the electron drift velocity, consists of only a dc
part ${\bf v}_0$ and a stationary time-dependent part ${\bf v}(t)$ of the form
\begin{equation}
{\bf V}(t)={\bf v}_0+{\bf v}_1 \cos(\omega t)+{\bf v}_2 \sin(\omega t).
\end{equation} 
 On the other hand, in order to carry out the statistical average we need the 
density matrix $\hat{\rho}$. For 2D systems having 
electron sheet density of order of, or higher than, 10$^{15}$ m$^{-2}$, 
the intrasubband and intersubband Coulomb interaction are sufficiently strong 
that it is adequate to describe the relative-electron transport state using a single 
electron temperature $T_{\rm e}$. Hence, the density matrix can be obtained from  
solving the Liouville equation by starting from the initial state of the
relative-electron-phonon system at time $t=-\infty$, in which the phonon system is 
in equilibrium at lattice temperature $T$ and the relative-electron
system is in equilibrium at an electron temperature $T_{\rm e}$:
\begin{equation}
\hat{\rho}|_{t=-\infty}=\hat{\rho}_0=\frac 1Z {\rm e}^{-H_{\rm er}/T_{\rm e}}{\rm e}^{-H_{\rm ph}/T}
\end{equation}
$Z$ is the normalized parameter.

With the density matrix thus obtained to the first order in $H_{\rm ei}+H_{\rm ep}$,
we can carry out the statistical average of operator equations
(\ref{eqdotv}) and (\ref{eqopdh}).
In the procedure the form factor 
\begin{eqnarray}
A({\bf q},t,t^{\prime})&\equiv& 
{\rm e}^{-{\rm i}{\bf q}\cdot \int_{t^{\prime }}^{t}{\bf v}%
(s)ds}
=\sum_{n=-\infty }^{\infty }{\rm J}_{n}^{2}(\xi ){\rm e}^{{\rm i}({\bf q}\cdot {\bf v}_0-n\omega) (t-t^{\prime
})}\nonumber\\
&&+\sum_{m\neq 0}{\rm e}^{{\rm i}m(\omega t-\varphi )}\left[ \sum_{n=-\infty }^{\infty
}{\rm J}_{n}(\xi ){\rm J}_{n-m}(\xi ){\rm e}^{{\rm i}({\bf q}\cdot {\bf v}_0-n\omega) (t-t^{\prime })}\right]
\end{eqnarray}
is expanded in terms of the Bessel functions ${\rm J}_n(x)$.
Here $\xi\equiv \sqrt{({\bf q}_\|\cdot {\bf v}_1)^2+({\bf q}_\|\cdot {\bf v}_2)^2}/\omega$
and $\varphi$ satisfies the relation
$\tan \varphi=({\bf q}\cdot {\bf v}_2)/({\bf q}\cdot {\bf v}_1)$.

Since all the transport properties are measured over a time interval much longer than
the period of terahertz field and we are concerned only with photoresponse and 
photoabsorption of the system, it suffices for us to know the frictional-force 
for the time oscillating term with base frequency $\omega$ and the energy-related 
quantity for the time-averaged term. The frictional force can be written as
\begin{equation}
{\bf F}(t)= {\bf F}_0-({\bf F}_{11}-{\bf F}_{22})\sin(\omega t)+({\bf F}_{12}+{\bf F}_{21})\cos(\omega t),
\end{equation}
with the functions ${\bf F}_0$ and ${\bf F}_{\mu \nu},(\mu,\nu=1,2)$ given by
\begin{equation}
{\bf F}_{0} =\sum_{{\bf q}_\|}{\bf q}_\|\left| U({\bf q}_\|%
)\right| ^{2}%
\sum_{n=-\infty }^{\infty }{\rm J}_{n}^{2}(\xi )\Pi _{2}({\bf %
q}_\|,\omega_0-n\omega )+ \nonumber
\sum_{{\bf q},\lambda }{\bf q}_\|\left| M({\bf q},\lambda )\right|
^{2}\sum_{n=-\infty
}^{\infty }{\rm J}_{n}^{2}(\xi )\Lambda _{2}({\bf q},\lambda, \omega_0+\Omega _{{\bf q}\lambda }-n\omega ),
 \label{eqf0}
\end{equation}
\begin{equation}
{\bf F}_{1\mu} =-\sum _{{\bf q}_\|}{\bf q}_\|\eta_{\mu}\left| U({\bf q}_\|%
)\right| ^{2}\sum_{n=-%
\infty }^{\infty }\left[ {\rm J}_{n}^{2}(\xi )\right] ^{\prime }\Pi _{1}(%
{\bf q}_\|,\omega_0-n\omega )- 
\sum_{{\bf q},\lambda }{\bf q}_\|\eta_{\mu}\left| M({\bf q},\lambda )\right|
^{2}\sum_{n=-\infty
}^{\infty }\left[ {\rm J}_{n}^{2}(\xi )\right] ^{\prime }\Lambda _{1}({\bf q%
},\lambda, \omega_0+\Omega _{{\bf q}\lambda }-n\omega ),\label{eqf1}
\end{equation}
\begin{equation}
{\bf F}_{2\mu} =\sum_{{\bf q}_\|}{\bf q}_\|\frac{\eta_{\mu}}{\xi}\left| U({\bf q}_\|%
)\right| ^{2}%
\sum_{n=-\infty }^{\infty }2n{\rm J}_{n}^{2}(\xi )\Pi _{2}({\bf %
q}_\|,\omega_0-n\omega )+ 
\sum_{{\bf q},\lambda }{\bf q}_\|\frac{\eta_{\mu}}{\xi}\left| M({\bf q},\lambda )\right|
^{2}\sum_{n=-\infty
}^{\infty }2n{\rm J}_{n}^{2}(\xi )\Lambda _{2}({\bf q},\lambda, \omega_0+\Omega _{{\bf q}\lambda }-n\omega ).
 \label{eqf2}
\end{equation}
In these expressions,
$\eta_{\mu}\equiv {\bf q}_\|\cdot {\bf v}_{\mu}/\omega \xi$;
$\omega_0\equiv {\bf q}_\|\cdot {\bf v}_0$;
$U({\bf q}_\|)$ is related to the impurity potential $U({\bf q}_\|,z_a)$
and distribution of impurities along $z$ axis $n_i(z)$;\cite{Lei851}
$\Lambda_2({\bf q},\lambda,\Omega)$
is the imaginary part of the electron-phonon correlation
function, which can be expressed through the imaginary part of electron density
correlation function $\Pi_2({\bf q}_\|,\Omega)$ as \cite{Lei3}
\begin{equation}
\Lambda_2({\bf q},\lambda,\Omega)=2\Pi_2({\bf q}_\|,\Omega)\left [n\left(\frac
{\Omega_{{\bf q}\lambda}}T\right)-
n\left(\frac{\Omega}{T_{\rm e}}\right)\right ],
\end{equation}
with $n(x)\equiv 1/[\exp(x)-1]$ being the Bose function.
The real parts of the electron-phonon correlation function
and the electron density correlation function, $\Lambda_1({\bf q},\lambda,\Omega)$ and
$\Pi_1({\bf q}_\|,\Omega)$, 
can be obtained from their imaginary parts through Kramers-Kronig transformation.\cite{Lei3}

The momentum-balance equation obtained by taking the statistical average
of operator equation (\ref{eqdotv}), has the following form
\begin{equation}
{\bf v}_1 \omega \sin(\omega t)-{\bf v}_2 \omega \cos(\omega t)=
\frac{1}{N_{\rm e}m}{\bf F}(t)+\frac{e}{m} \left\{{\bf E}_0+{\bf E}(t)+
[{\bf v}_0+{\bf v}(t)] \times {\bf B}\right\}.
\label{eqnbase1}
\end{equation} 
That is
\begin{eqnarray}
0&=&N_{\rm e}e{\bf E}_{0}+N_{\rm e} e ({\bf v}_0 \times {\bf B})+
{\bf F}_0,\label{eqv0}\\
{\bf v}_{1}&=&\frac{e{\bf E}_s}{m\omega}-\frac{1}{N_{\rm e}m\omega }
\left( {\bf F}_{11}-{\bf F}_{22}\right)
-\frac{e}{m\omega }({\bf v}_{2}\times
{\bf B}),\label{eqv1}\\
-{\bf v}_{2}&=&\frac{e{\bf E}_c}{m\omega}+\frac{1}{N_{\rm e}m\omega }
\left( {\bf F}_{12}+{\bf F}_{21}\right) -\frac{e}{m\omega }({\bf v}_{1}
\times {\bf B}).\label{eqv2}
\end{eqnarray}
The energy-balance equation is obtained by taking the long-time average
of statistically averaged operator equation (\ref{eqopdh}) to be
\begin{equation}
N_{\rm e}e{\bf E}_0\cdot {\bf v}_0+S_{\rm p}- W=0.
\label{eqnbase2}
\end{equation}
Here $W$ is the time-averaged rate of the energy transfer from the electron system 
to the phonon system, whose expression can be obtained from the second term
on the right side of equation (\ref{eqf0}) by replacing the ${\bf q}_\|$
factor with $\Omega_{{\bf q}\lambda}$.
$S_{\rm p}$ is the time-averaged rate of the electron energy gain from the radiation
field and has the following form
\begin{equation}
S_{\rm p} =\sum_{{\bf q_\|}}\left| U({\bf q}_\|%
)\right| ^{2}\sum_{n=-\infty }^{\infty }n\omega {\rm J}_{n}^{2}(\xi )\Pi _{2}(%
{\bf q_\|},\omega_{0}-n\omega )+
\sum_{{\bf q,}\lambda }\left| M({\bf q,}\lambda )\right|
^{2}\sum_{n=-\infty }^{\infty }n\omega {\rm J}_{n}^{2}(\xi )\Lambda _{2}(%
{\bf q,}\lambda, \omega_{0}+\Omega _{{\bf q}\lambda
}-n\omega ).
\end{equation}
Note that $S_{\rm p}$ is negatively equal to
time averaged Joule heat $\left <{\bf v}(t)\cdot{\bf F}(t)\right >_{t}=
N_{\rm e}e({\bf E}_{0} \cdot {\bf v}_0 +{\bf E_{s}\cdot{\bf v}_2}/2
+{\bf E_{c}\cdot{\bf v}_1}/2)$.

Momentum and energy balance equations (\ref{eqv0}) to (\ref{eqnbase2}) constitute a close
set of equations to determine the parameters ${\bf v}_0$, ${\bf v}_1$,
${\bf v}_2$, and $T_{\rm e}$ when ${\bf E}_0$, ${\bf E}_c$ and ${\bf
E}_s$ are given.

The sum over $n$ in the expressions for ${\bf F}_0$, ${\bf F}_{\mu\nu}$, $W$ and 
$S_{\rm p}$ represents contribution of all orders of multiphoton processes related to 
the photons of frequency $\omega$. In the present formulation, the role of the 
single-frequency radiation field is two fold. (1) It induces photon-assisted impurity 
and phonon scatterings associated with single ($|n|=1$) and multiple $(|n|\ge 1)$
photon processes, which are superposed on the direct impurity and phonon scattering 
($n=0$) term. (2) It transfers energy to the electron system ($S_{\rm p}$) through
single and multiple photon-assisted process.   

Note that $\Pi_{2}({\bf q}_{\|},\Omega)$ and $\Pi_{1}({\bf q}_{\|},\Omega)$ are 
respectively the imaginary and real parts of the electron density correlation
function of the 2D system in the presence of the magnetic field. 
In the Landau representation, one can write\cite{Ting}
\begin{eqnarray}
\Pi _2({\bf q}_{\|},\Omega ) & = & \frac 1{2\pi
l^2}\sum_{n,n'}C_{n,n'}(l^2q_{\|}^2/2)\Pi _2(n,n',\Omega),\\
\label{pi_2}\Pi _2(n,n',\Omega)&=&-\frac2\pi \int d\varepsilon
\left [ f(\varepsilon )- f(\varepsilon +\Omega)\right ]
{\rm Im}G^r_n(\varepsilon +\Omega){\rm Im}G^r_{n'}(\varepsilon ),
\end{eqnarray}
where $l=\sqrt{1/|eB|}$ is the magnetic length,
\begin{equation}
C_{n,n+l}(Y)=\frac{n!}{(n+l)!}Y^le^{-Y}[L_n^l(Y)]^2
\end{equation}
with $L_n^l(Y)$ being the associate Laguerre polynomial, $f(\varepsilon
)=\{\exp [(\varepsilon -\mu)/T_{\rm e}]+1\}^{-1}$ is the Fermi distribution
function, and ${\rm Im}G_n^r(\varepsilon )$ is the imaginary part of the
Green's function of the Landau level $n$,
which is proportional to the density of
states, such that the density of electrons is given by
\begin{equation}
\label{den}N_{\rm e}=-\frac 1{\pi ^2l^2}\sum\limits_n\int d\varepsilon
f(\varepsilon ){\rm Im}G^r_n(\varepsilon ). 
\end{equation}
This equation determines the chemical potential.

In principle, to obtain the Green's function 
of Landau levels $n$, ${\rm G}^r_n(t)$,
a self-consistent calculation has to be carried out from the Dyson equation for
the self-energy with all the scattering mechanisms included.\cite{Leadley}
The resultant Green's function is generally a complicated function of
the magnetic field, temperature, and Landau-level index $n$, also
dependent on the relative strength of the impurity and phonon scattering.
In the present study we do not attempt a self-consistent calculation of
${\rm G}^r_n(t)$. Instead, we choose a Gaussian-type function for the
Landau-level shape for simplicity,\cite{Gerhardts1}
\begin{equation}
{\rm G}^r_n(t)=-i\Theta(t)\exp[-i(n-\frac 12)\omega_ct-\frac 12 \Gamma _n ^2 t^2]
\end{equation}
with a unified broadening parameter $\Gamma _n=\Gamma$ for all the Landau
levels, which is taken as $(2e\omega _c /\pi m \mu_0)^{1/2}$.
When the hot-electron effect is neglected,
$\mu_0$ corresponds to the linear mobility at temperature $T$
in the absence of magnetic fields.\cite{Ando}
In order to consider the hot-electron-induced Landau levels
broadening, we will empirically treat $\mu _0$
as the linear mobility of the system in the absence of the magnetic field
at temperature $T_{\rm e}$. Note that
this Gaussian-type of Green's
function is proved to be correct at low temperature.\cite{Gerhardts2} On the other hand, it also
has been used to interpret the magnetophonon resonance at high lattice temperature.\cite{XGWu3}
In the present paper we will show that this Green's function can lead to
a qualitative agreement between theoretical and experimental results 
within the magnetic field range considered.
To improve the agreement further, a more careful study on the Green's function of a Landau level
should be performed.\cite{Leadley}

Above formulation can be used to describe the transport and optical properties
of magnetically-biased quasi-2D semiconductors subjected to a dc field and a terahertz 
field. The conventional magneto-optical study
in the far-infrared frequency regime
corresponds to the case of zero dc field ${\bf E}_0=0$,
where one studies the intensity-dependent terahertz absorption, transmission 
and other effects in the presence of a strong magnetic field. On the other hand,
to investigate photoconductivity,
we should treat the weak dc field limit of our formulation.

\section{Cyclotron resonance in drift velocity}

Substituting the force Eq.\,(\ref{eqv1}) into Eq.\,(\ref{eqv2}), we can write
\begin{equation}
{\bf v}_{1}=(1-{\omega_c^2}/{\omega^2})^{-1}\left\{
\frac{e}{m\omega}\left[{\bf E}_s+\frac{e}{m\omega }(
{\bf E}_c\times{\bf B})\right]
-\frac{1}{N_{\rm e}m\omega }
\left( {\bf F}_{11}-{\bf F}_{22}-
\frac{e}{m\omega }
\left[({\bf F}_{12}+{\bf F}_{21})\times
{\bf B}\right]\right)\right\},\label{vv1}
\end{equation}
\begin{equation}
-{\bf v}_{2}=(1-{\omega_c^2}/{\omega^2})^{-1}\left \{
\frac{e}{m\omega}\left[{\bf E}_c-\frac{e}{m\omega }(
{\bf E}_s\times{\bf B})\right]
+\frac{1}{N_{\rm e}m\omega }
\left( {\bf F}_{12}+{\bf F}_{21}+
\frac{e}{m\omega }
\left [ ({\bf F}_{11}-{\bf F}_{22})\times
{\bf B}\right]\right)\right\}.\label{vv2}
\end{equation}
Cyclotron resonance is easily seen in the case of weak scatterings when 
the terms with ${\bf F}_{\mu \nu}$ functions in the above equation are small: 
both ${\bf v}_1$ and ${\bf v}_2$ exhibit peaks at CR.
Since all the transport quantities, including $W$, $S_{\rm p}$ and ${\bf F}_{0}$, 
are functions of the drift velocity as well as the electron temperature $T_{\rm e}$, 
and the latter is determined by the energy balance equation (30),
the CR of ${\bf v}_1$ and ${\bf v}_2$ will
result in the CR in $W$, $S_{\rm p}$, ${\bf F}_{0}$ and $T_{\rm e}$.

 Eqs. (\ref{vv1}) and (\ref{vv2}) can be further simplified when the radiation field is
weak and the dc field is absent. In this case ${\bf v}_{\mu}$
can be treated as small parameters. To the first order of these
small parameters, the force function ${\bf F}_{\mu\nu}$ can be written as
\begin{equation}
{\bf F}_{\mu\nu}=N_{\rm e}m{\bf v}_{\nu}M_{\mu}(\omega,{\bf v}_0)
\,\,\,\,\,\,\,\,\,\,(\mu,\nu=1,2), \label{f12}
\end{equation}   
where $M_{\mu}(\omega,{\bf v}_0)$ are the real ($\mu=1$)
and imaginary ($\mu=2$)
parts of the memory functions.\cite{Lei3}
It is convenient to write out the expression for the complex velocity
$v_+\equiv v_{1x}+{\rm i} v_{2x}$ and $v_- \equiv v_{2y}-{\rm i}
v_{1y}$ rather than for ${\bf v}_1$ and ${\bf v}_2$
\begin{eqnarray}
v_+&=&e \tau /2m^*\left(\frac {E_+}{(\omega-\omega_c^*)\tau+{\rm i}}+
\frac {E_-}{(\omega+\omega_c^*)\tau+{\rm i}}\right),\nonumber \\
v_-&=&-e \tau /2m^*\left(\frac {E_+}{(\omega-\omega_c^*)\tau+{\rm i}}-
\frac {E_-}{(\omega+\omega_c^*)\tau+{\rm i}}\right).\label{barv}
\end{eqnarray} 
Here, we have defined
\begin{eqnarray}
m^*&=&m[1+M_1(\omega,{\bf v}_0)/\omega],\nonumber\\
1/\tau&=&M_2(\omega,{\bf v}_0)/[1+M_1(\omega,{\bf v}_0)/\omega],\nonumber\\
\omega_c^*&=&eB/m^*,\nonumber\\
E_+&=&E_{sx}+E_{cy}+{\rm i}(E_{sy}-E_{cx}),\nonumber\\
E_-&=&E_{sx}-E_{cy}-{\rm i}(E_{sy}+E_{cx}),\nonumber
\end{eqnarray} 
with $(E_{sx},E_{sy})\equiv {\bf E}_{s}$ and
$(E_{cx},E_{cy})\equiv {\bf E}_{c}$.

Our weak-field results for ${\bf v}_1$ and ${\bf v}_2$ reduces to those of 
Ref.\,\onlinecite{Ting} in the case of circularly polarized ac fields.

\section{Transmission}

For normally incident electromagnetic wave, the transmitted electric
field ${\bf E}(t)$, which is regarded as the field driving the 2D electrons,\cite{Chiu}
is related to the incident electric field ${\bf E}_{\rm i}(t)$ by
\begin{equation}
{\bf E}(t)=\frac{N_{\rm e}e{\bf v}(t)/\epsilon_0c}{n_{\rm
s}+n_0}+\frac{2n_0}{n_{\rm s}+n_0}{\bf E}_{\rm i}(t).
\label{eqet}
\end{equation} 
Here $n_0$ and $n_{\rm s}$ are the relative refractive indices of
the airs and 2D semiconductors, and $c$ and $\epsilon_0$ are the light
speed and the dielectric constant in vacuum, respectively. The transmitted field 
${\bf E}(t)$ depends on the drift velocity
${\bf v}(t)$ of the 2D system. In the following numerical studies on transmission and
photoconductivity, we will assume a sinusoidal incident field ${\bf E}_{\rm i}(t)
=(E_{{\rm i}s}\sin(\omega t), 0)$ along $x$ axis 
and derive ${\bf E}(t)$ self-consistently together with ${\bf v}(t)$.

We have numerically calculated the magneto-optical properties
of a GaAs/AlGaAs heterojunction subjected to a THz ac field and a
magnetic field. The strength of transmitted ac field ${\bf E}(t)$, and
the parameters ${\bf v}_\mu$($\mu$=1,\,2) and
$T_{\rm e}$, are obtained
by resolving the Eqs.\,({\ref{eqet}),
(\ref{eqv1}), (\ref{eqv2}) and (\ref{eqnbase2}).
We consider a GaAs-based quasi-2D system having electron density $N_{\rm e}=2.5\times 10^{15}$\,m$^{-2}$ and 4.2\,K linear mobility
50\,m$^2$/Vs (which is used to determine the impurity density)
at lattice temperature $T=$4.2\,K,
similar to that used in Ref.\,\onlinecite{Rodriguez}. 
The elastic scattering due to randomly
distributed charged impurity and
the inelastic scattering due to polar optical phonons (via Fr\'ohlich coupling
with electrons), longitudinal acoustic phonons (via deformation potential and
piezoelectric coupling), and transverse acoustic phonons (via piezoelectric coupling with
electrons) are taken into account.
The material and electron-phonon coupling parameters are taken as typical
values for GaAs. In the numerical calculation the maximum Landau level is 
taken to be 20, and the summation over multiphoton indices $n$ are carried up to
a given accuracy of $10^{-3}$ for each quantity.

The transmittance ${\cal T}$, defined as\cite{Chiu}
\begin{equation}
{\cal T}=\frac {<|{\bf E}(t)|^2>_t}{<|{\bf E}_{\rm i}(t)|^2>_t}
\end{equation}
with $<..>_t$ denoting the time average. When connecting it with measured quantities, of
course, the multiple interference between interfaces of the substrate has to be taken into
acount. The calculated transmittance and the corresponding electron temperature
are plotted in Fig.\,1 as functions of the intensity
of the THz field at two frequencies $\omega/2\pi=0.83$ and 1.6\,THz 
in the center position of CR, namely $\omega_c = \omega$.
It can be seen that, the transmittance first decreases gently with increasing intensity of the 
THz radiation from zero, and reaches a bottom at a critical intensity around 
10\,W/cm$^2$, then increases rapidly with further increasing the field strength. 
This feature appears more pronounced at lower frequency, in consistence
with the experimental observation,\cite{Rodriguez} as shown in the inset of Fig.\,1,
where the measured transmittance for 1.6\,THz and 0.24\,THz exhibits similar trend.
The case of 0.83\,THz does not show minimum.
This deviation is believed to come from the experimental errors.
In Fig.\,2 we display the transmittance CR line shape for incident
electromagnetic fields of different intensities at frequency 0.83\,THz.
The line width exhibits no significant change below the critical intensity but
increases rapidly when the intensity of THz 
field grows above the critical value.

This kind of $E_{{\rm i}s}$-dependent behavior of transmittance is in agreement
with the intensity dependence of the absorption rate 
$\alpha\sim S_{\rm p}/E_{{\rm i}s}^{2}$ in the absence of the magnetic field.
For a similar 2D GaAs-based semiconductor without magnetic field, 
Ref.\,\onlinecite{Lei00} showed that,
the absorption percentage increases with increasing strength of the
radiation field from low-field value, then reaches a maximum (of order of 2 percent) 
at the field amplitude  
around several kV/cm before decreasing with further increase of the radiation field 
strength, and that lower frequency has stronger maximum. At low velocity side, 
when hot-electron effect is relatively weak and the direct impurity and phonon 
scatterings change little, the behavior of the absorption rate comes mainly from 
the drift-velocity dependence of the multi-photon assisted scattering matrix element, 
as described by the Bessel functions ${\rm J}_n^2(\xi)$ in the expressions for 
$S_{\rm p}$.  In fact,  
all the multi-photon ($n\ge 1$) contributions to the absorption coefficient are zero 
at vanishing velocity and reach maxima at finite (increasing with $n$) 
drift velocities, the resultant absorption coefficient first increases with
increasing velocity. When the drift velocity becomes sufficiently large, 
reduction of absorption rates, induced by the large argument of the lowest-order Bessel
functions, will exceed the increased
contributions from the other multi-photon processes. This 
leads to the drop of the absorption rate.
In the present case having a strong magnetic field, 
CR greatly enhances the drift velocity  
${\bf v}_1$ and ${\bf v}_2$ at $\omega_c\sim \omega$ for a given $E_{{\rm i}s}$ 
in comparison with the case without magnetic field. Therefore, the maximum 
absorption rate should appear at much smaller strength
of radiation field and have much larger value for the case of cyclotron resonance 
than in the absence of a magnetic field or far away from CR. 

The behavior of the transmittance CR line shape is related to Landau level broadening 
due to hot-electron effect. For THz field below the critical intensity the electron 
temperature is less than 60\,K (see Fig.\,1) and impurities are the dominant 
scatterers, yielding
almost a constant mobility $\mu_0$ (thus the Landau level broadening). When the THz
field goes above the critical intensity, the electron temperature grows rapidly and 
polar optical phonons become the dominant scatterers, giving rise to a strongly 
temperature-dependent mobility $\mu_0$, thus a Landau level broadening which increases
rapidly with increasing field strength.

Our formulation can also be employed to investigate
the change of the transmitted electromagnetic field polarization
in quasi-two-dimensional electron systems.
This effect is known as Faraday effect and
has long been investigated under linear condition.\cite{Faraday,Connel,Volkov}
The present approach provides an convenient formulation to study the Faraday effect 
for the case when the incident light is strong and the nonlinear
absorption occurs.

The relevant quantities characterizing the Faraday effects are
the ellipticity $\eta$ and
Faraday rotation angle $\theta_{\rm F}$, which are determined
through the amplitudes of transmitted field ${\bf E}(t)$
\begin{eqnarray}
\tan \eta&=&(a^+-a^-)/(a^++a^-),\\
\theta_{\rm F}&=&(\phi^+-\phi^-)/2.
\end{eqnarray}
Where 
\begin{eqnarray}
\tan\phi^+&=&\frac{E_{sx}+E_{cy}}{E_{cx}-E_{sy}},\\
\tan\phi^-&=&\frac{-E_{sx}+E_{cy}}{E_{cx}+E_{sy}},\\
a^+&=&\sqrt{(E_{sx}+E_{cy})^2+(E_{cx}-E_{sy})^2}/2,\\
a^-&=&\sqrt{(E_{sx}-E_{cy})^2+(E_{cx}+E_{sy})^2}/2,
\end{eqnarray}   
with $(E_{sx},E_{sy})\equiv {\bf E}_{s}$ and
$(E_{cx},E_{cy})\equiv {\bf E}_{c}$.

We plot the calculated results of $\eta$ and $\theta_{\rm F}$ in Fig.\,3 
for the above mentioned two-dimensional sample.
The resonance in ellipticity and antiresonance in Faraday angle
can be seen evidently. Their line shapes also manifest different behavior when
the intensity of THz field lies below or above the critical value.
These intensity-dependent behaviors of ellipticity and Faraday
rotation can also be understood by multi-photon-assisted scatterings 
and hot-electron effect induced Landau level broadening.

\section{Photoconductivity}

The response of the linear dc conductance to far-infrared irradiation is easily
obtained in the the weak dc field limit of our formulation.
Taking ${\bf v}_0$ to be in the $x$ direction,
${\bf v}_0=(v_{0x},0,0)$ and expanding the equation (\ref{eqv0}) 
to the first order in $v_{0x}$, we obtain the
transverse and longitudinal resistivities $R_{xy}$ and  $R_{xx}$ as follows:
\begin{eqnarray}
R_{xy}\equiv\frac {E_{0y}}{N_{\rm e}ev_{0x}}&=&B/N_{\rm e}e,\\
R_{xx}\equiv\frac {E_{0x}}{N_{\rm e}ev_{0x}}&=&-\frac 1{N_{\rm e}^2 e^2}\sum_{{\bf q}_\|}q_x^2\left|
U({\bf q}_\|)\right| ^2\sum_{n=-\infty }^\infty {\rm J}_n^2(\xi)\left[ \frac \partial {\partial\, \Omega }\Pi
_2({\bf q}_\|,\Omega )\right] _{\Omega =-n\omega }\nonumber\\
&&- \frac 1{N_{\rm e}^2e^2}
\sum_{ {\bf q},\lambda } q_x^2\left| M ( {\bf
q},\lambda )\right| ^2\sum_{n=-\infty }^\infty {\rm J}_n^2(\xi)\left[ \frac \partial {\partial\, \Omega }\Lambda
_2({\bf q},\lambda,\Omega )\right]_{\Omega =\Omega_{{\bf q}\lambda}-n\omega}.\label{rxx}
\end{eqnarray}
The parameters ${\bf v}_1$, ${\bf v}_2$ and $T_{\rm e}$ in these
expressions should be determined by solving equations (\ref{eqv1}), (\ref{eqv2})
and (\ref{eqnbase2}) with zero ${\bf v}_0$. The longitudinal photoresistivity
is defined as
\begin{equation}
\Delta R_{xx}\equiv R_{xx}-R_{xx}^{0},
\end{equation}
with $R_{xx}^{0}$ being the longitudinal magnetoresistivity in the absence of
the radiation field.

Photoconductivity in semiconductors, in the absence or in the presence of magnetic fields,
has long been known at low temperatures, and was understood to result from the effects of 
electron heating due to the absorption of the radiation field energy.\cite{New,Kogan,Russ2}
In our formulation, the photoconductivity arises not only from
the hot-electron effect (electron temperature change), but also from the
photon-assisted electron-impurity and electron-phonon scatterings.
Although it is difficult to distinguish contributions to photoconductivity
from different mechanisms when the applied terahertz field is strong,
in the case of weak ac fields, the longitudinal
photoresistivity can be written as the sum of two terms:
\begin{equation}
\Delta R_{xx}=\Delta R_{xx}^{({\rm h})}+\Delta R_{xx}^{({\rm op})}.
\end{equation}
The first term $\Delta R_{xx}^{({\rm h})}$ is obtained through
expanding Eq.\,(\ref{rxx}) by the small parameter
$\Delta T_{\rm e}=T_{\rm e}-T$ and
is the result of ac field
induced electron temperature change, as that proposed first by Kogan for
the case without a magnetic field.\cite{Kogan}
After determining the small electron temperature change from
energy-balance equation (\ref{eqnbase2}), we can write
\begin{equation}
\Delta R_{xx}^{({\rm h})}=\Phi\cdot\Delta T_{\rm e}.
\end{equation}
Here
\begin{equation}
\Phi\equiv\left(\frac {\partial R_{xx}^{0}}
{\partial T_{\rm e}}\right)_{T_{\rm e}=T}
=\frac {\partial}{\partial T} R_{xx}^{0}-
\frac{2}{N_{\rm e}^2e^2}\sum_{{\bf q},\lambda }q_x^2\left| M({\bf q},\lambda )\right|
^{2}\frac{\Omega _{{\bf q}\lambda }}{T^2}n^{\prime }( \frac{\Omega _{{\bf q}\lambda}}{T} )
\left[\frac {\partial}{\partial \Omega}\Pi _{2}({\bf
q}_\|,\Omega)\right]_{\Omega=\Omega _{{\bf q}\lambda }},
\end{equation}
and
\begin{equation}
\Delta T_{\rm e} =
\Xi\cdot\left [8\sum_{{\bf q},\lambda}\frac{\Omega_{{\bf q}\lambda}^2}{T^2}
\left| M({\bf q},\lambda )\right|^{2}\Pi _{2}({\bf q%
}_\|,\Omega_{{\bf q}\lambda})n^{\prime }( \frac{\Omega _{{\bf
q}\lambda}}{T} )\right]^{-1},
\end{equation}
with
\begin{equation}
\Xi=\frac 1{\omega^2}({\bf v}_1^2+{\bf v}_2^2)
\left[N_{\rm e}m\omega M_2(\omega,0)-
\sum_{{\bf q},\lambda,\pm}\Omega_{{\bf q}\lambda}{\bf q}_\|^2
\left| M({\bf q},\lambda )\right|^{2}\Lambda_{2}({\bf q%
}_\|,\Omega_{{\bf q}\lambda}\pm \omega)\right ].\label{eqxi}
\end{equation}
Note that the terms on the right hand side of Eq.\,(\ref{eqxi}) are 
respectively the changes of $S_{\rm p}$ and $W$ induced by photon-assisted scatterings.

The second component of photoresistivity $\Delta R_{xx}^{({\rm op})}$ is the result of
the photon-assisted scattering processes and ac field induced electron
distribution change
\begin{equation}
\Delta R_{xx}^{({\rm op})}=\frac 1{4N_{\rm e}^2{\rm e}^2\omega}
(3v_{1x}^2+3v_{2x}^2+v_{1y}^2+v_{2y}^2)
Q_2(\omega).
\end{equation}
Where 
\begin{equation}
Q_2(\omega)=\sum_{q_\|}q_\|^4\left [ A_{d}({\bf q}_\|,0)-A_{d}({\bf
q}_\|,\omega)\right ]/\omega,
\end{equation}
with
\begin{equation}
A_d({\bf q}_\|,\Omega)=\left| U({\bf q}_\|%
)\right| ^{2} \frac{\partial}{\partial \Omega}\Pi _{2}(%
{\bf q}_\|,\Omega )+
\sum_{q_z,\lambda}\left| M({\bf q},\lambda )\right|
^{2}\frac{\partial}{\partial \Omega}\Lambda _{2}({\bf q%
},\lambda,\Omega).\label{psixx}
\end{equation}

In the weak ac field limit we can see from Eq.\,(\ref{barv}) that the
amplitudes of time-dependent drift velocity are linearly dependent
on the strength of applied THz fields. Consequently, the
photoresistivity is proportional to the intensity of THz field.
When the intensity of the driving field becomes strong, the dependence of
photoconductivity on the strength of THz field exhibits
a complicated behavior. At the same time, the contribution from
the electron-temperature change and nonthermal photon-assisted scattering are hybridized.
At low lattice temperatures, the hot-electron effect is sufficiently strong and is
generally the dominant mechanism for photoresistivity. At high temperature, however, 
when the polar optical phonon scattering provides
an efficient energy dissipation channel, the photoconductivity
is mainly contributed from nonthermal mechanism.

Recently, the photoconductivity CR at high temperature has been
demonstrated in the experiment of Ref. \,\onlinecite{Koenraad}.
The remarkable peaks in photoconductivity are observed when the cyclotron 
frequency is closed to the frequency of terahertz fields. It
is also found that, not only the height but also the width of CR peaks
increases with increasing intensity of terahertz fields.

In order to illustrate CR in photoconductivity at high lattice
temperatures, we have numerically evaluated the dc longitudinal
photoresistivity of a magnetically-biased GaAs/AlGaAs heterojunction.
The lattice temperature $T$ is $150$\,K.
The considered sample has electron
density $N_{\rm e}=2.0\times 10^{15}$\,m$^{-2}$ and 4.2\,K linear mobility
200\,m$^2$/Vs, similar to that used in the experiment of Ref. \,\onlinecite{Koenraad}.

In Fig.\,4 the longitudinal photoresistivity induced by THz fields of frequency 
$\omega/2\pi=4\,$THz having several different amplitudes is plotted
as a function of magnetic field. The resonant structure
near the cyclotron resonance position shows up clearly. Furthermore, 
with increasing strength of THz field, the CR peaks ascend and the line shapes broaden.
These features are in qualitative agreement with experimental results
in Ref.\,\onlinecite{Koenraad}.
In the inset of Fig.\,4 we plot the photoresistivity
at CR as a function of the intensity of THz field. One can see that
the photoresistivity follows a linear dependence on the intensity of THz field 
in the range $0<I_{{\rm i}s}<1$\,kW/cm$^2$. For larger ac field intensity,
the deviation from linear dependence appears.
The electron temperature also exhibits resonant peak when the cyclotron frequency 
is closed to the THz frequency, as shown in Fig.\,5. 
Nevertheless, since at high lattice temperatures the strong electron-LO phonon
scattering provides an efficient energy dissipation channel, the rise of electron
temperature is modest, and
the hot-electron effect induced photoconductivity
is small in comparison with the nonthermal effect. The
photon-assisted scatterings are the main mechanisms for
the photoconductivity at high temperature. For relative weak radiation field, e.g.
$E_{is}\le 0.24$\,kV/cm, the electron temperature has no appreciable difference
from $T$, yet the $\Delta R_{xx}$ still exhibits sizable resonance, comparable with
the peak height observed in experiments of Ref. \onlinecite{Koenraad}
In Fig.\,6 we plot the longitudinal
photoresistivity as a function of magnetic field strength for
relatively weak ac fields.
The contribution to photoconductivity from
hot-electron effect is less than 3\,\%. We also show the experimental observation\cite{Koenraad}
of CR in photoconductivity in the inset of Fig.\,6.
The extraordinary width of the resonance shown in the experimental $\Delta R_{xx}$ is possible
due to the short pulse or large bandwidth of the terahertz radiation used in experiment,
as explained in Ref. \onlinecite{Koenraad}.

\section{conclusion}

We have developed the momentum and energy balance equations for steady-state electron
transport and optical absorption under the influence of a dc electric field, 
an intense THz ac electric field in a two-dimensional 
semiconductor in the presence of a strong magnetic field perpendicular to 
the 2D plane.  This formulation allows us to investigate the THz-field-intensity 
dependence of the cyclotron resonance in transmittance and photoconductivity of
GaAs/AlGaAs heterojunctions. We found that the CR peaks and line shapes of
transmittance exhibit different behaviors when the intensity of
the THz field increases in the range above or below a certain critical value.  
The cyclotron resonance in photoresistivity, however, always enhances with increasing
intensity of the THz field. These results qualitatively agree with the
experimental observations. The intensity-dependent behavior of
transmittance at CR is explained as the result of
combing hot-electron effect induced Landau level broadening and 
electron-phonon scattering enhancement, and the drift velocity dependence of the 
photon-assisted scattering matrix elements.
We have also clarified that the CR in photoconductivity
is not only the result of the electron heating, but also comes from photon-assisted 
scattering enhancement, especially at high temperatures. 
The effect of an intense THz field on Faraday angle and
ellipticity of magnetically-biased 2D semiconductor systems have also been
demonstrated.

The authors gratefully acknowledge stimulating discussions with
Drs. Bing Dong and W.S. Liu.
This work was supported by the National Science Foundation of China 
(Grant Nos.\,60076011 and 90103027), the Special Funds
for Major State Basic Research Project (Grant No.\,20000683), 
the Shanghai Municipal Commission of Science and Technology, and
the Shanghai Postdoctoral Fellow Science Foundation.

\vspace{0.5cm}
\centerline{Figure Captions}
\vspace{0.5cm}

FIG.\,1. The intensity-dependence of 2D semiconductor transmittance and electron
temperature,
is plotted at cyclotron resonance
position, $\omega_c=\omega$.
The transmittance is normalized to the value of zero magnetic field.
The THz fields with two different frequency $\omega/2\pi=0.83$ and
$\omega/2\pi=1.6$\,THz
are exposed to the studied 2D system. The lattice temperature is $T=4.2$\,K.
Experimental results of transmittance versus the intensity of THz fields
$I_{{\rm i}s}$ (Fig.\,2 of Ref. \onlinecite{Rodriguez}) is reproduced in the inset.
\vspace{0.5cm}

FIG.\,2. The cyclotron resonance of 2D semiconductor transmittance
of several incident electromagnetic
fields with a same frequency $\omega/2\pi=0.83$\,THz but different intensity
$I_{{\rm i}s}=0.1$,\,$0.6$,\,$74$,\,$750$,\,$1500$,\,$2500$\,W/cm$^2$.
\vspace{0.5cm}

FIG.\,3. The Faraday angle $\theta_{\rm F}$ and ellipticity $\eta$
are plotted as functions of
the strength of magnetic fields $B$ for the same system and under
the same condition as described in Fig.\,2.
\vspace{0.5cm}

FIG.\,4. The cyclotron resonance in the longitudinal resistivity
change $\Delta R_{xx}$ induced by a radiation field of frequency $\omega/2\pi=4$\,THz
having several different amplitudes
$E_{{\rm i}s}=0.24$,\,$0.61$,\,$0.87$,\,$1.1$,\,$1.2$\,kV/cm. The lattice
temperature is $T=150$\,K. The inset shows $\Delta R_{xx}$ versus intensity of
incident THz fields $I_{{\rm i}s}$ at $\omega_c=\omega$.
\vspace{0.5cm}

FIG.\,5. The cyclotron resonance in electron temperature $T_{\rm e}$
for the same system and under
the same condition as described in Fig.\,4.
\vspace{0.5cm}

FIG.\,6.  Similar to the Fig.\,4, but the strength of THz fields are
relatively weak  $E_{{\rm
i}s}=0$,\,$0.086$,\,$0.13$,\,$0.19$,\,$0.24$\,kV/cm. The experimental
results (Fig.\,3 of Ref. \onlinecite{Koenraad}) of the cyclotron resonance in photoconductivity is
shown in the inset.

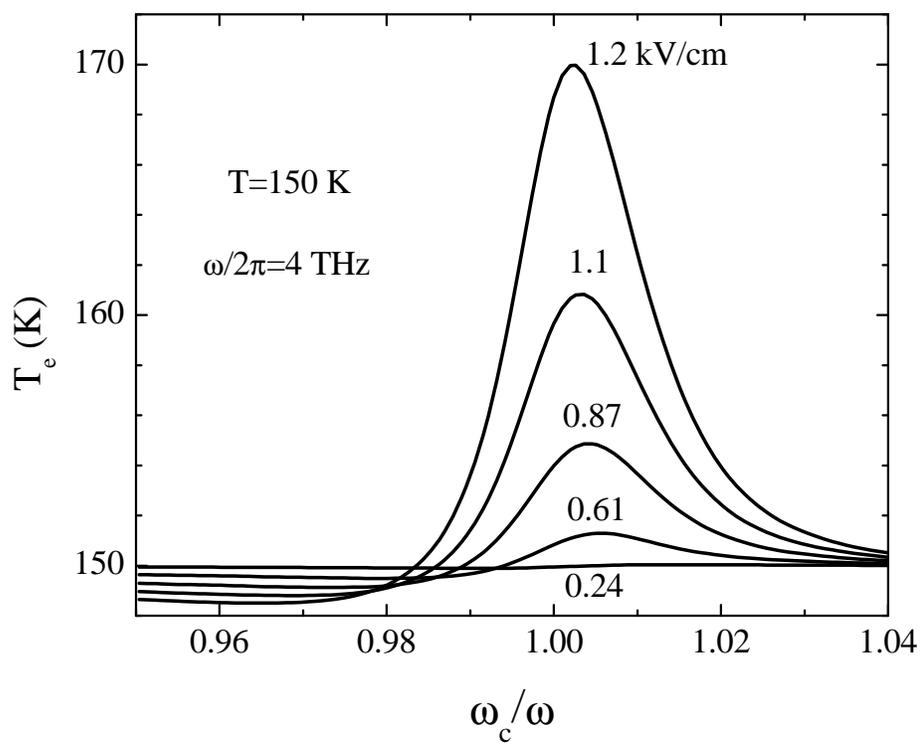
\end{document}